\input phyzzx.tex
\tolerance=1000
\voffset=-0.0cm
\hoffset=0.7cm
\sequentialequations
\def\rl{\rightline}

\def\t1{{\tilde 1}}

\def\t{\theta}

\REF{\GUT}{A. H. Guth, Phys. Rev {\bf D23} (1981) 347.}
\REF{\LIN}{A. D. Linde, Phys. Lett. {\bf B108} (1982) 389.}
\REF{\ALB}{A. Albrecht and P. J. Steinhardt, Phys. Rev. Lett. {\bf 48} (1982) 1220.} 
\REF{\DSS}{G. Dvali, Q. Shafi and S. Solganik, hep-th/0105203.}
\REF{\BUR}{C. P. Burgess at. al. JHEP {\bf 07} (2001) 047, hep-th/0105204.}
\REF{\ALE}{S. H. Alexander, Phys. Rev {\bf D65} (2002) 023507, hep-th/0105032.}
\REF{\EDI}{E. Halyo, hep-ph/0105341.}
%\REF{\SHI}{G. Shiu and S.-H. H. Tye, Phys. Lett. {\bf B516} (2001) 421, hep-th/0106274.}
\REF{\CAR}{C. Herdeiro, S. Hirano and R. Kallosh, JHEP {\bf 0112} (2001) 027, hep-th/0110271.}
\REF{\SHA}{B. S. Kyae and Q. Shafi, Phys. Lett. {\bf B526} (2002) 379, hep-ph/0111101.}
\REF{\BEL}{J. Garcia-Bellido, R. Rabadan and F. Zamora, JHEP {\bf 01} (2002) 036, hep-th/0112147.}
\REF{\KAL}{K. Dasgupta, C. Herdeiro, S. Hirano and R. Kallosh, Phys. Rev. {\bf D65} (2002) 126002, hep-th/0203019.}
\REF{\TYE}{N. Jones, H. Stoica and S. H. Tye, JHEP {\bf 0207} (2002) 051, hep-th/0203163.}
\REF{\OTH}{T. Matsua, hep-th/0302035; hep-th/0302078; T. Sato, hep-th/0304237; Y. Piao, X. Zhang and Y. Zhang, hep-th/0305171; C. P. Burgess, 
J. Cline and R. Holman, hep-th/0306079.}
\REF{\LAS}{E. Halyo, hep-th/0307223.}
%\REF{\CHA}{A. D. Linde, Phys. Lett. {\bf B129} (1983) 177; Phys. Lett. {\bf B175} (1986) 395.}
%\REF{\HYB}{A. D. Linde, Phys. Lett. {\bf B259} (1991) 38; Phys. Rev. {\bf D49} (1994) 748.}
\REF{\FRA}{M. Bertolini, hep-th/0303160 and references therein.} 
\REF{\DTE}{E. Halyo, Phys. Lett. {\bf B387} (1996) 43, hep-ph/9606423.} 
\REF{\BIN}{P. Binetruy and G. Dvali, Phys. Lett. {\bf B450} (1996) 241, hep-ph/9606342.}
\REF{\TYP}{E. Halyo, Phys. Lett. {\bf B454} (1999) 223, hep-ph/9901302.}
\REF{\BRA}{E. Halyo, Phys. Lett. {\bf B461} (1999) 109, hep-ph/9905244; JHEP {\bf 9909} (1999) 012, hep-ph/9907223.}
\REF{\MAP}{C. L. Bennett et. al., astro-ph/0302207; G. Hinshaw et. al., astro-ph/0302217; A. Kogut et. al., astro-ph/0302213.}
\REF{\JAU}{D. E. Diaconescu, M. R. Douglas and J. Gomis, JHEP {\bf 9802} (1998) 013, hep-th/9712230.}
\REF{\EGU}{T. Eguchi and A. J. Hanson, Ann. Phys. {\bf 120} (1979) 82.}
\REF{\DGM}{M. R. Douglas, B. R. Greene and D. R. Morrison, Nucl. Phys. {\bf B506} (1997) 84, hep-th/9704151.}
\REF{\QUI}{M. R. Douglas and G. W. Moore, hep-th/9603167; C. V. Johnson and R. C. Myers, hep-th/9610140.}
\REF{\GIB}{G. W. Gibbons and S. W. Hawking, Comm. Math. Phys. {\bf 66} (1979) 291.}
\REF{\QUI}{B. Ratra and P. J. E. Peebles, Phys. Rev. {\bf D37} (1988) 3406.}
\REF{\STE}{L. Wang, R. R. Caldwell, J. P. Ostriker and P. J. Steinhardt, Astrophys. {\bf J530} (2000) 17, astro-ph/9901388.}
\REF{\TEV}{E. Halyo, JHEP {\bf 0110} (2001) 025, hep-th/0105216.}
%\REF{\STR}{A. Strominger, JHEP {\bf 0110} (22001) 034; hep-th/0106113.}
%\REF{\INF}{A. Strominger, hep-th/0110087.}
%\REF{\HOL}{E. Halyo, hep-th/0203235.}

\singlespace
%\rl{SU-ITP-01-03}
\rl{hep-th/0312042}
\rl{\today}
%\rl{T}
\pagenumber=0
\normalspace
\medskip
\bigskip
\titlestyle{\bf{Inflation on Fractional Branes: D--Brane Inflation as D--term Inflation}}
\smallskip
\author{ Edi Halyo{\footnote*{e--mail address: vhalyo@stanford.edu}}}
\smallskip
%\centerline {Department of Physics} 
%\centerline{Stanford University} 
%\centerline {Stanford, CA 94305}
%\centerline{and}
 \centerline{California Institute for Physics and Astrophysics}
\centerline{366 Cambridge St.}
\centerline{Palo Alto, CA 94306}
\smallskip
\vskip 2 cm
\titlestyle{\bf ABSTRACT}

We describe a D--brane inflation model which consists of two fractional $D3$ branes separated on a transverse $T^2 \times K3$. Inflation arises due to the resolved 
orbifold singularity of $K3$ which corresponds to an anomalous D--term on the brane. We show that D--brane inflation in the bulk corresponds to D--term inflation on the brane. The inflaton 
and the trigger field parametrize the interbrane distances on $T^2$ and $K3$ respectively. After inflation the branes reach a supersymmetric configuration in which they are at the origin 
of $T^2$ but separated along the $K3$ directions.

\singlespace
\vskip 0.5cm
\endpage
\normalspace

\centerline{\bf 1. Introduction}
\medskip

The most promising way to obtain inflation[\GUT-\ALB] in string theory seems to be by D--brane inflation[\DSS]. In this scenario, inflation occurs on D--branes due to their slow motion
in the bulk which arises from weak interbrane potentials. Generically, the inflaton is the distance or the angle between two D--branes. There have been many attempts to construct models of
D--brane inflation[\DSS-\OTH]. Unfortunately, some of the most interesting models do not (and probably cannot) have compactified extra transverse dimensions 
and therefore are not realistic[\CAR,\BEL]. Other models with compactified extra dimensions are quite complicated and not easy to analyze[\KAL].

In this letter, we construct a very simple model of D--brane inflation on fractional $D3$ branes[\FRA]. We consider a $D3$ brane tansverse to $K3 \times T^2$ where the $K3$ 
is smooth. However, consider the $K3$ first
to be in its orbifold limit, $T^4/Z_2$. Near one of the 16 fixed points of $K3$, the relevant geometry that the $D3$ brane feels is the ALE space $R^4/Z_2$. The $Z_2$ moding requires the
presence of an image $D3$ brane with $R^4$ coordinates opposite to those of the original $D3$ brane. As long as the brane and its image are away from the orbifold fixed plane they
cannot move on the $T^2$. On the other hand, if the brane and its image are stuck at the fixed point they can 
move independently on the transverse $T^2$. They then become fractional D--branes with half the charge and tension of a regular $D3$ brane.
This state is supersymmetric; in fact it belongs to the Coulomb branch of the field theory on the fractional branes. Therefore, there is no potential between the two fractional branes.
We can break supersymmetry and lift the Coulomb branch by blowing up the orbifold singularity. We simply replace the singularity by a small sphere ($P^1$) whose radius is a
modulus. As a result, there is a potential between the two fractional $D3$ branes and they start moving slowly towards each other on the $T^2$. This motion corresponds to inflation on 
(either one of) the fractional
branes. Inflation ends when the branes are close enough so that the slow--roll conditions are violated. While the branes approach each other on the $T^2$ they start to
separate in the $T^4/Z_2$ directions (because the fields that describe the motion in these directions become tachyonic). The final configuration is supersymmetric and given by the branes 
at the same point in $T^2$ (origin of the Coulomb branch) but separated on the $T^4/Z_2$ (on the Higgs branch). 

On the fractional branes, the above scenario is described by D--term inflation. The resolution of the orbifold singularity corresponds to the introduction of an anomalous 
D--term on the brane world--volume where
the magnitude of the anomalous D--term is fixed by the radius of the blown up $P^1$. The interbrane distance along the $T^2$ directions corresponds the the inflaton whereas that along the
$T^4/Z_2$ directions corresponds to the trigger field. In this letter, we analyze the above model on the brane world--volume and show that D--brane inflation in the bulk
corresponds to D--term inflation[\DTE,\BIN] on the brane. Due to the
${\cal N}=2$ supersymmetry of the system, the brane world--volume description is equivalent to the bulk description of two fractional branes approaching each other. 
Since the branes can move only 
on the $T^2$ the bulk potential is logarithmic which agrees with the logarithmic potential generated in D--term inflation due to one--loop effects.

The main advantages of our D--brane inflation model are its simplicity and the fact that it has compact transverse dimensions. These are due to the absence of higher dimensional 
branes in the model such as the $D6$ brane that is present
in Hanany--Witten configurations[\CAR,\LAS]. These higher dimensional branes make simple compactifications very difficult if not impossible. Models with $D7$-$D3$ branes[\KAL] can be 
compactified but
they are quite cumbersome due to conditions arising from the presence of the $D7$ branes. In our model we have only $D3$ branes because we can break supersymmetry 
and obtain inflation by deforming the background, i.e. blowing up the orbifold singularity. This relation between the resolution of a singularity in the background geometry and the
origin of inflation is very intriguing.

This letter is organized as follows. In section 2, we biefly review D--term inflation. In section 3, we describe our brane inflation model in terms the fractional $D3$ brane 
world--volume theory and show that it corresponds to D--term inflation. The last section includes a discussion of our results and our conclusions.

\bigskip
\centerline{\bf 2. D--term Inflation}
\medskip

In this section we briefly review D--term inflation[\DTE-\BRA] which requies two fields: the inflaton and the trigger field. The generic
potential is
$$V(\phi_1,\phi_2,\Phi)=\lambda^2(|\phi_1|^2+|\phi_2|^2) |\Phi|^2+ \lambda^2 |\phi_1|^2 |\phi_2|^2+ g^2 (|\phi_1|^2-|\phi_2|^2 - \xi)^2 \eqno(1)$$
Here the inflaton and the trigger field are $\Phi$ and $\phi_1$ respectively. In addition to the Yukawa terms (with coupling $\lambda$) the potential contains an anomalous 
D--term $\xi$ which is
the origin of inflation. Note that there is no inflaton mass at tree level; a small nonzero inflaton mass arises from one--loop effects and drives inflation.

Now let us assume that the initial conditions are given by $\Phi>>\xi>\phi_1, \phi_2$. Then $\phi_{1,2}$ have large masses and settle very quickly to the minimum of their potential at
$\phi_1=\phi_2=0$. In this configuration, supersymmetry is broken due to the anomalous D--term. As a result, there is a one--loop potential
$$V_{1-loop}=g^2 \xi^2 \left(1+{g^2 \over {16 \pi^2}} log{\lambda \Phi \over M_P} \right) \eqno(2)$$
which leads to a small inflaton mass
$$m^2_{\Phi}=\xi^2{g^4 \over {16 \pi^2}} {1 \over {\Phi^2}} \eqno(3)$$
Due to this small mass, the inflaton rolls down its potential slowly which results in inflation. Inflation ends when $\Phi$ reaches a critical value $\Phi=\Phi_c$
at which the slow--roll conditions are violated. Later, when $\Phi^2<\Phi_c^2=2g^2 \xi/\lambda^2$, the trigger field becomes tachyonic; $\phi_1=0$ becomes a local maximum and the trigger
field starts rolling down to its new minimum at $\phi_1^2=\xi$. The final state of the system is given by $\Phi=0$ and $\phi_1=\sqrt \xi$ which is supersymmetric.

In order to be realistic, our model of D--brane inflation has to satisfy the WMAP constraints[\MAP]. First and foremost, the slow--roll constraints given by $0<\epsilon_1<0.022$ and 
$-0.06<\epsilon_2<0.05$ have to be satisfied. Here $\epsilon_1=\epsilon$
and $\epsilon_2=2(\epsilon-\eta)$ which are defined by the slow--roll parameters
$$\epsilon={M_P^2 \over 2} \left(V^{\prime} \over V \right)^2 \eqno(4)$$
and
$$\eta=M_P^2 \left(V^{\prime \prime} \over V \right) \eqno(5)$$
Inflation occurs when $\epsilon,\eta <<1$ and ends when at least one of them becomes of $O(1)$.
Our model also has to produce the correct amount of scalar density perturbations; $18.8\times 10^{-10}<A_s<24.8 \times 10^{-10}$ where
$$A_s={H^2 \over {8 \pi^2 M_P^2 \epsilon_1}} \eqno(6)$$
$A_s$ is related to the magnitude of density perturbations by $\delta \rho/\rho \sim A_s^2$.
The ratio of the amplitudes for the tensor and scalar perturbations must satisfy $0<R<0.35$ where $R \sim 16 \epsilon_1$. The scalar and tensor spectral indices which parametrize
the deviations from scale invariant perturbations are constrained by
$0.94<n_s<1.02$ and $-0.044<n_t<0$ where
$$n_s \sim 1-2 \epsilon_1 -\epsilon_2 \qquad \qquad n_t \sim -2\epsilon_2 \eqno(7)$$
In addition, the models must result in about 60 e--folds of inflation
$$N = M_P^{-2} \int {V \over V^{\prime}} d\phi \sim 60 \eqno(8)$$

Using the total potential given by eqs. (1) and (2) we find for the slow--roll conditions
$$\epsilon_1={g^4 \over {2(16 \pi^2)^2}}{M_P^2 \over \Phi^2} \qquad \qquad \epsilon_2={{2g^2} \over {(16 \pi^2)^2}}{M_P^2 \over \Phi^2}(1+g^2) \eqno(9)$$
The number of e--folds is given by
$$N={{16 \pi^2} \over {g^2M_P^2}}(\Phi_i^2-\Phi_c^2) \eqno(10)$$ 
where $\Phi_i,\Phi_c$ are the initial and critical values of the inflaton field respectively.
The magnitude of scalar density perturbations is
$$A_s={{64 \pi^2} \over {g^2}} {{\xi^2 \Phi^2} \over {M_P^6}} \eqno(11)$$

\bigskip
\centerline{\bf 3. D--Brane Inflation as D--term Inflation on Fractional Branes}
\medskip

In this section, we describe our D--brane inflation model which realizes inflation on fractional $D3$ branes. We consider a $D3$ brane (along directions $0,1,2,3$) transverse to 
$K3 \times T^2$ where 
the $K3$ is taken to be at its orbifold limit, i.e $T^4/Z_2$. We assume that the $D3$ brane is
close to the orbifold singularity and therefore the space that it sees is $ALE \times T^2$ where the ALE space is given by the orbifold $R^4/Z_2$.
We take the $T^2$ and $R^4$ to be along the $4,5$ and $6,7,8,9$ directions respectively. We also define for simplicity the complex planes $s=X_4+iX_5$, $u=X_6+iX_7$ and $v=X_8+iX_9$.
Due to the $Z_2$ moding, the $D3$ brane at any point on the ALE space has an image with the opposite $R^4$ coordinates. In addition to moving on $R^4$, the brane (and its image) can be
stuck at the fixed point of the orbifold, $u=v=0$. In this case, the brane and its image are free to move independently on the $T^2$ along the $s$ direction. It is well--known that 
then we get two fractional $D3$ branes with
half the charge and tension of a regular $D3$ brane[\FRA]. These factional branes can also be described as $D5$ branes wrapped around a two sphere ($P^1$) with zero radius[\JAU]. 
The nonzero tension of the fractional branes is due to the nonzero flux of the $B$ field through this vanishing cycle.
$$T_{D3}={1 \over {(2 \pi)^6 g_s \ell_s^6}} \int_{S^2} \sqrt{det B}={1 \over {2 (2 \pi)^4 g_s \ell_s^4}} \eqno(12)$$
where we used the fact that the B--flux through the vanishing sphere is $(2 \pi \ell_s)^2/2$.
The configuration that we described above is supersymmetric and therefore there is no force between the fractional branes. However, supersymmetry can be broken by resolving (blowing up) 
the orbifold singularity i.e. by replacing the singular point at $u=v=0$ by a sphere ($P^1$) with nonzero radius. As a result, the background becomes a smooth Eguchi--Hanson space.
In this new background, there is an attractive potential between the fractional branes which leads to inflation.

We now analyze the above scenario on the world--volume of one of the fractional $D3$ branes. The field theory on a $D3$ brane transverse to $R^4/Z_2 \times T^2$ is given by an ${\cal N}=2$
supersymmetric $U(1)_1 \times U(1)_2$ gauge theory with two hypermultiplets $\phi_{1,2}$ with charges $(1,-1)$ and $(-1,1)$ respectively[\FRA]. The gauge fields arise from the excitations
of strings with both ends on the same (fractional) brane in the $X_0,X_1,X_2,X_3$ directions. The scalar fields in the gauge multiplets arise from the excitations of
the same strings along the $X_4,X_5$ directions. The hypermultiplets arise from the excitations of strings which connect the brane with its 
image in the $X_6,X_7,X_8,X_9$ directions. One combination of $U(1)$'s gives the center
of mass motion and is not interesting. Thus, we will consider the $U(1)$ gauge group (and the corresponding scalar) which is given by the difference between the two original ones, 
$(1/2)[U(1)_1-U(1)_2]$. Under this group
the hypermltiplets have charges of $1$ and $-1$ respectively. $\phi_{1,2}$ together with the scalar coming from the gauge multiplet (which we denote by $\Phi$) describe the 
relative positions of the $D3$ brane and its image in the transverse directions. In particular, $\Phi$ gives the relative position of the branes on the $s$ plane ($4,5$ directions) 
whereas $\phi_{1,2}$ parametrize the relative position on the $u$ and $v$ planes ($6,7$ and $8,9$ directions) respectively. We can write
$$\Phi={{X_4+iX_5} \over {2 \pi \ell_s^2}} \qquad \phi_1={{X_6+iX_7} \over {2 \pi \ell_s^2}} \qquad \phi_2={{X_8+iX_9} \over {2 \pi \ell_s^2}} \eqno(13)$$
The superpotential is fixed by the ${\cal N}=2$ supersymmetry to be
$$W(\Phi,\phi_1,\phi_2)=g \Phi \phi_1 \phi_2 \eqno(14)$$
Note that, due to the ${\cal N}=2$ supersymmetry the Yukawa coupling is equal to the gauge coupling given by 
$${1 \over g^2}={e^{-\phi} \over {(2 \pi)^3 \ell_s^2}}  \int_{S^2}  \sqrt{det(G+B)}={1 \over {4 \pi g_s}} \eqno(15)$$
where $\phi$ is the space--time dilaton and $S^2$ is the sphere with vanishing radius at the orbifold fixed point.
Together with the D--term contribution the above superpotential leads to the scalar potential
$$V(\phi_1,\phi_2,\Phi)=g^2(|\phi_1|^2+|\phi_2|^2) |\Phi|^2+ g^2 |\phi_1|^2 |\phi_2|^2+ g^2 (|\phi_1|^2-|\phi_2|^2 )^2 \eqno(16)$$
We see from the above potential that the moduli space (or the space of brane configurations) has two branches: the Coulomb branch with $\phi_{1,2}=0$ and $\Phi \not =0$ in which
the fractional branes are separated in the $s$ direction and the Higgs branch with $\Phi=0$ and $\phi_1, \phi_2 \not=0$ in which the branes are separated in the $u,v$ directions.

Now let us take the two fractional branes to be on the Coulomb branch, i.e. at $u=v=0$ but with $s\not =0$. We note that this is not strictly necessary as an initial condition.
If the initial state is one in which the two branes are separated by a large distance on both $T^2$ and $K3$, the branes will reach the fixed point $u=v=0$ very fast which agrees with
the evolution in D--term inflation.
We can blow up the orbifold singularity at $u=v=0$ by replacing this point
with a two sphere of finite radius, $R$. When the ALE orbifold singularity is replaced by a $P^1$ the geometry is described by the Eguchi--Hanson space[\EGU] with the metric
$$ds^2=\left(1-\left({R \over r}\right)^4 \right)^{-1} dr^2+ r^2 \left(1-\left({R \over r}\right)^4 \right)(d \psi+cos \theta d\phi)^2+r^2(d\theta^2+sin^2\theta d\phi^2) \eqno(17)$$
This is a smooth ALE space with a global $Z_2$ identification (just like $R^2/Z_2$). The point $r=R$ is a coordinate singularity if the angle $\psi$ has a period of $2 \pi$ rather than
the more natural $4 \pi$. Near $R$, the space looks like $R^2_{r \psi} \times S^2_{\theta \phi}$. The $S^2$ is of radius $R$ and corresponds exactly to the blown up sphere that
replaced the orbifold fixed point.

The resolution of the singularity breaks supersymmetry and lifts the Coulomb branch. The only remaining supersymmetric configuration is the one with both branes at $s=0$ and separated
along the $K3$ directions.
This means that the fractional branes feel an attractive potential. In our model, from the bulk point of view, the resolution of the singularity is the origin of brane inflation.
The two fractional branes move slowly towards each other which results in D--brane inflation. After inflation ends, the final state is given by the branes at the origin of the $s$
plane but separated in the $u$ direction, i.e. on the Higgs branch.

In the brane world--volume theory, blowing up the orbifold singularity at $u=v=0$ corresponds to the addition of an anomalous D--term $\xi$ to the scalar potential in eq. (16)[\DGM]. 
This is easy to see since the presence of this term eliminates the origin of the Higgs branch $\phi_1=\phi_2=0$ (which corresponds to the fixed point) as required.
To be precise, due to the ${\cal N}=2$ supersymmetry the resolution of the singularity gives three anomalous D--terms, $\xi_{1,2,3}$ which form an $SU(2)$ triplet. However, 
we can always choose the anomalous D--term to be one particular component by an $SU(2)$ rotation.
Thus the potential now becomes
$$V(\phi_1,\phi_2,\Phi)=g^2(|\phi_1|^2+|\phi_2|^2) |\Phi|^2+ g^2 |\phi_1|^2 |\phi_2|^2+ g^2 (|\phi_1|^2-|\phi_2|^2 -\xi)^2 \eqno(18)$$
which is the potential for D--term inflation. Due to supersymmetry breaking during inflation, there is a one--loop potential for the inflaton which is logarithmic as in 
eq. (2). This is exactly
what one would expect from the bulk point of view since the fractional $D3$ branes can move only in two dimensions. In fact, ${\cal N}=2$ supersymmetry
guarantees that the bulk result agrees with the one obtained on the world--volume. The reason is string channel duality for highly supersymmtric (i.e. {${\cal N}=2$) configurations
which states that one loop open string (gauge theory) results must agree with the tree level closed string (gravity) results.

In order to complete the bulk--brane dictionary given by eqs. (13) and (15) we need to express the anomalous D--term
in terms of the string variables. This can be done by equating the extra energy due to $\xi$ in the scalar potential to the energy of a $D5$ brane wrapped on the blown--up orbifold
singularity with radius $R$. (We remind that the fractional $D3$ brane can be seen as a $D5$ brane wrapped on the vanishing sphere.) Thus we have
$$g^2 \xi^2=T_{D5} A_{S^2}= {{4 \pi R^2} \over {(2 \pi)^6 g_s \ell_s^6}} \eqno(19)$$
which gives
$$\xi= {1 \over {4 \pi^2 g_s}}{R \over \ell_s^3} \eqno(20)$$ 

Inflation ends when the the branes are close enough so that the slow--roll conditions are violated, i.e. $\epsilon \sim 1$ or $\eta \sim 1$. Later, when the branes are closer than
$s<\sqrt{2 R \ell_s/g_s}$, the trigger field $\phi_1$ becomes tachyonic. This means that there is an instability in the $u$ direction. The branes start to separate on the $u$ plane 
until they reach the minimum of their potential at $\phi_1=\sqrt \xi$. Note that in our model the tachyonic instability causes a separation of the two fractional branes rather than their 
collapse on top each other. As a result, the final state is a very simple one, i.e. two branes separated by a distance of $\sqrt{R \ell_s/g_s}$ in the $u$ direction as opposed 
to branes within branes.

Using the bulk--brane dictionary given by eqs. (13), (15) and (20) and $M_P^2=L^6/g_s^2 \ell_s^8$ where $L$ denotes the common radius of $T^2 \times K3$ directions
we can obtain the WMAP constraints in terms of the string parameters. The strongest constraint on the initial 
distance between the branes (in the $s$ direction) arises from eq. (8) which gives $\Phi_i/M_P \sim 0.3$ (assuming $g_s \sim 1/2$) or $s_i \ell_s^2/L^3 \sim 1$.
Using this relation we find that $\epsilon_1 \sim 10^{-5}$ and $\epsilon_2 \sim 10^{-4}$ which are very small. This is a generic result in D--term inflation due to the one--loop
inflaton potential which is proportional to $(g^2/16 \pi^2)$. From eq. (7) we find that the scalar and tensor spectral indeces are $n_s =1$ and $n_t=0$ up to $O(10^{-4})$. 
Thus the model predicts scalar and tensor perturbations which are very nearly scale invariant.
The magnitude of the anomalous D--term or the blow up radius $R$ is fixed by the correct amount of scalar
density perturbations. From eqs. (11) and (20) we get $\xi/M_P^2 \sim 10^{-11}$ or $R \ell_s^2/L^3 \sim 10^{-5}$.

\bigskip

\centerline{\bf 4. Conclusions and Discussion}

\medskip

In this letter, we described a D--brane inflation model on fractional branes. We considered a $D3$ brane and its image on $T^2 \times K3$. When the branes are stuck at one of the 
fixed points of 
$K3$ they can move independently on the $T^2$. Supersymmetry of this configuration (the Coulomb branch) can be broken by resolving the singularity and replacing the orbifold
by an Eguchi--Hanson space. This results in an attractive logarithmic potential between the branes which move slowly towards each other on the $T^2$. Inflation corresponds to this
motion. On the brane world--volume, this scenario is described by D--term inflation in which the inflaton and the trigger field correspond to the interbrane distances on the $T^2$ and
$K3$ respectively. On the brane, there is a one--loop inflaton potential which is logarithmic. This agrees with the bulk description of D--brane inflation since the branes can only move
in two dimensions (on the $T^2$). We also obtained a complete bulk--brane dictionary and showed that the string parameters can satisfy the WMAP constraints. 

Our analysis of the model was based completely on the brane world--volume theory, i.e. D--term inflation. As we noted, due to the ${\cal N}=2$ supersymmetry bulk results must agree with 
those on the brane. However, the direct bulk calculation of the interbrane potential is quite complicated even though the evolution of the two branes in the bulk is conceptually clear.
Usually, using supergravity, the potential is obtained by substituting the metric and dilaton generated by one brane into the world--volume action of the other one. In our case,
this is not easy to do since we need to include the Eguchi-Hanson metric of the background. Alternatively, a one--loop open string calculation between the two branes s complicated by the
fact that the space between the branes is not flat. It would be nice if the bulk interbrane potential can be obtained explicitly and shown to agree with our results.

The advantage of our model is its simplicity and the fact that it has compact transverse dimensions. As mentioned in the introduction, these are related to the fact that the
model does not contain any higher dimensional branes. We note that the our pair of branes is not intrinsically unstable like $D$-${\bar D}$ pairs which have to annihilate at the
end of inflation. In D--brane inflation models with two branes of different dimensions, at the end of inflation the branes in question end up in the Higgs phase where they become fluxes, 
instantons etc. In our model, the final state is much simpler given by the two branes separated by a finite distance in the $v$ direction.
The origin of inflation in our model is the resolution of the orbifold singularity which is an effect of the background geometry. 
This is to be contrasted with other models in which the origin of inflation is a small deviation from a supersymmetric brane configuration such as branes at small angles or with small fluxes.

We considered above only the simplest orbifold singularity possible, i.e. $T^4/Z_2$. The orbifold group can be generalized to any $Z_n$ or maybe to more complicated orbifolds. In the 
$T^4/Z_n$ case, there will be $n$ fractional branes falling towards each other. D--brane inflation along the lines of our model can arise on any one of them. The world--volume
field theory on these fractional branes is a quiver theory[\QUI], i.e $U(1)^n$ with hypermultiplets in the bi--fundamental representations. Resolving the singularity 
replaces the orbifold with a gravitational instanton or Gibbons--Hawking space[\GIB]. On the branes, this corresponds to the addition of an anomalous D--term for each one of the $U(1)$'s. 
D--brane inflation can be obtained by considering any two of the adjacent $U(1)$'s which effectively reproduces our model.  

Any model of inflation can also be considered a model of quintessence[\QUI,\STE] for a different range of parameters. In our case, we would obtain quintessence on a D--brane[\TEV] if
$g^2 \xi \sim 10^{-60}M_P^2$ or using eqs. (15) and (20) $R \ell_s^5/L^6 \sim 10^{-60}$. This can be satisfied by an extemely small blow--up radius $R \sim 10^{-60} \ell_s$ with
$\ell_s^{-1} \sim 10^{18}~ GeV$. Of course, this is extreme fine--tuning which is expected from any configuration that accomodates the observed nonzero cosmological constant.

%\bigskip

%\centerline{\bf Acknowledgements}

\vfill

\refout

\end
\bye